\documentclass[11pt]{article}
\usepackage{latexsym}
\newcommand{\beq}{\begin{equation}} 
\newcommand{\eeq}{\end{equation}}
\newcommand{\beqs}{\begin{eqnarray}} 
\newcommand{\eeqs}{\end{eqnarray}}
\newcommand{\tr}{\mathrm{tr}}
\newcommand{\WW}{{\cal W}}
\newcommand{\refe}[1]{(\ref{#1})}
\begin{document}
\begin{center}
{\LARGE \bf Konishi Anomaly}\\
\smallskip
{\LARGE \bf and Chiral Gauge Theories\footnote{
To appear in the proceedings of the 27th Johns Hopkins Workshop
``Symmetries \& Mysteries of M-Theory", G\"oteborg, August 24--26, 2003,  
and the RTN Workshop ``The quantum structure 
of spacetime and the geometric nature of fundamental interactions'',
Copenhagen, September 15--20, 2003.}}
\\
\vspace{1cm}
{\large
Riccardo Argurio}
\vskip 0.7cm
{\large \it  Physique Th\'eorique et Math\'ematique \\
\smallskip
Universit\'e Libre de Bruxelles, C.P. 231, 1050 Bruxelles, Belgium}
\vskip 0.3cm
\end{center}
\vspace{1cm}
\begin{abstract}
We begin with a brief introduction on ${\cal N}=1$ gauge theories, 
focusing on the importance of the effective superpotential in light of 
the new techniques to compute it systematically. We then proceed to consider 
theories for which the Konishi anomaly proves to be enough
to solve exactly for the effective superpotential. As an example we study
a chiral $SO(10)$ gauge theory, where we also
discuss the occurrence of dynamical supersymmetry breaking.
\end{abstract}

\vspace{1cm}

Gauge theories with minimal ${\cal N}=1$ supersymmetry (SUSY) are extensively
studied because, while they are believed to display QCD-like properties 
like confinement, dynamical generation of a mass gap and chiral symmetry
breaking, they are still more tractable due to some key SUSY properties
like the perturbative non-renormalization theorem and the holomorphy
of the effective superpotential. In particular, the SUSY features make it
possible to derive some exact results about the vacuum structure of these
theories. 

The crucial ingredient for deriving such results is, as said above, the
holomorphy of the superpotential, both in the fields (possibily effective)
and in the couplings \cite{naturalness}.
The perturbative non-renormalization theorem then implies that the 
(Wilsonian) effective superpotential is as follows:
\beq 
W_{eff}= W_{tree} + W_{non-pert},  \label{weff} 
\eeq
that is, only non-perturbative corrections are allowed to take place.
We will see that it is precisely these corrections that can be
computed in a systematic way. Once that $W_{eff}$ is known, the exact
SUSY vacua can be determined by extremizing it with respect to the
low energy effective (gauge invariant) fields that it depends on.
(This holds provided there are no singularities in the effective
K\"ahler metric, which are not expected in the cases we will consider
below.)

For SUSY gauge theories, one has to assume confinement and generation
of a mass gap in the gauge sector. The latter is implemented by the
introduction of the (RG invariant) holomorphic scale $\Lambda$, defined
in terms of the running gauge coupling $g(\mu)$ and the $\theta$ angle through:
\beq
\Lambda^\beta = \mu^\beta e^{-{8\pi^2 \over g^2(\mu)}+i\theta}, \label{scale}
\eeq
where $\beta$ is the (positive) coefficient of the one-loop-exact 
Wilsonian beta function.
The matter sector in its turn is described by a set of gauge invariant 
operators $X_r$ which at low energies become the effective fields.

It is then often possible to determine the effective superpotential in the
following way. For $W_{tree}=0$, it can only depend (holomorphically)
on the scale $\Lambda$ and on the effective fields $X_r$ and it becomes
possible to determine the form of $W_{eff}= W_{non-pert}(X_r,\Lambda)$ 
by symmetry arguments alone, and eventually fix the constant factor
by an instanton calculation. This approach is based on a case by case
analysis of different gauge theories. We will see below that there is a more
systematic approach to compute $W_{eff}$.

We can now add a tree level superpotential, which is typically
written in terms of the invariants $X_r$ together with their
associated couplings, $W_{tree}= \sum g_r X_r$.
The perturbative non-renormalization theorem and the requirement 
of good behaviour in several decoupling limits then imply 
what is often called the ``linearity principle'' \cite{integratein},
that is, the couplings $g_r$ only enter linearly in $W_{eff}$:
\beq 
W_{eff} = \sum g_r X_r +  W_{non-pert}(X_r,\Lambda),
\label{linea}
\eeq
namely, no dependence in $g_r$ is allowed in $W_{non-pert}$.

If $W_{tree}$ gives a mass to some or all of the matter fields, 
it makes sense to integrate out the massive $X_r$. This is done
by extremizing \refe{linea} with respect to $X_r$. The result
gives the v.e.v. of $X_r$ in terms of all the couplings and $\Lambda$
(assuming that all matter fields have been integrated out).
However, by writing the extremization equation as:
\beq
g_r = - {\partial W_{non-pert} \over \partial X_r}, \label{intout}
\eeq 
one sees that the linearity principle implies that integrating out $X_r$
is the same as performing a Legendre transform where $X_r$ and its coupling
$g_r$ are a conjugate pair.

One then obtains the effective superpotential in terms of the couplings
and the holomorphic scale, $W_{eff}(g_r, \Lambda)$. Thinking of integrating
out as a Legendre transform leads to invert the relation \refe{intout}
and integrate in the effective fields through:
\beq
X_r =  {\partial W_{eff} \over \partial g_r}. \label{intin}
\eeq
The linearity principle guarantees that
this procedure is exact as far as the superpotential of the (low energy)
effective theory is concerned. 
Of course, the relation \refe{intin} for the v.e.v. of a gauge invariant 
operator also follows from the path integral of the theory.

The ideas above extend to the gauge sector of the theory. Consider the
tree level action of the gauge theory, with the coupling running at one-loop.
We can write it as the following F-term:
\beq
-\beta S \log {\Lambda \over \mu}, \label{trees}
\eeq
where we have introduced the glueball superfield $S$, defined by:
\beq
 S = - \frac{1}{32 \pi^2} \tr \WW^\alpha \WW_\alpha, \qquad 
\WW_\alpha = -\sqrt{2}i \lambda_\alpha + \theta^\beta F_{\alpha\beta}+\dots.
\eeq
The term \refe{trees} can be interpreted as saying that $S$ and 
$-\beta \log \Lambda / \mu$ are conjugate in the same way as $X_r$ and $g_r$.
We can then trade the dependence on $\Lambda$ in $W_{eff}$ into dependence
on $S$ by simply integrating in $S$. We compute:
\beq
S= {1\over \beta} \Lambda {\partial W_{eff} \over \partial \Lambda},
\eeq 
and then invert it to reexpress $\Lambda = \Lambda(S, g_r)$.

Finally, we can express the effective superpotential in terms of this
new set of variables:
\beq
W_{eff}(S,g_r,\Lambda)= W_{VY}(S,\Lambda) + W_{pert}(S,g_r).
\label{wdv}
\eeq
We have split it into two parts. The first, $W_{VY}$, is the pure gauge
part and takes the Veneziano-Yankielowicz form \cite{VY}. (Note that we have
re-added the ``tree level'' term \refe{trees} so that the holomorphic
scale $\Lambda$ appears instead of the cut-off $\mu$.) By extremizing it
one gets, for instance for a pure $SU(N)$ theory, the relation
$S^N=\Lambda^{3N}$. 

The second part, $W_{pert}(S,g_r)$, can be computed systematically and
in a perturbative expansion, as pointed out recently in \cite{DV,proof,cdsw}. 
Note however that,
since by virtue of the linearity principle we do not lose any information
by integrating out and in again, exactly the same information
is contained in $W_{eff}(S,g_r)$ and in $W_{non-pert}(X_r,\Lambda)$.

The new techniques developped in \cite{DV,proof,cdsw} are reviewed 
in \cite{review}. Here we will instead focus on a class of theories
for which this approach is at best problematic. Indeed, the approach 
of \cite{cdsw} relies on the presence of a matter field in the adjoint
representation of the gauge group, while the perturbative method
of \cite{proof}, which applies to more generic representations, requires
the matter fields to have a mass. Both of these requirements are not met
by purely ${\cal N}=1$ theories (i.e. with no adjoint) with chiral matter 
content.\footnote{A chiral theory containing also adjoint matter has
been considered in \cite{llt}. In this case the techniques of \cite{cdsw}
are applicable.}

On the other hand, in purely  ${\cal N}=1$ gauge theories one can see
that the Konishi anomaly relations \cite{konishi} are often enough
to solve for $W_{pert}(S,g_r)$ -- see \cite{binor} for several
examples of this, including some particular chiral theories.

Here we will review how the Konishi anomaly can be used to
determine the effective superpotential, and then apply the method
to a chiral theory which displays dynamical SUSY breaking.

The Konishi anomaly for a $U(1)$ rotation of the superfield $\Phi$
can be roughly written as:
\beq
\bar D^2 \Phi^\dagger e^V \Phi = \sum \# g_r X_r -  \# S,
\label{koni}
\eeq
where on the LHS the action of $V$ is always in the representation
to which $\Phi$ belongs, 
and on the RHS the numerical coefficients depend on the specific
theory being considered. The first term is generated by the tree
level superpotential, while the second term is generated at one loop.
There will be as many relations as there are
fundamental matter fields.

The highest component of \refe{koni} is nothing else than the usual
chiral anomaly. More interesting to us is the lowest component
of \refe{koni}. It is a relation between chiral operators, i.e. precisely
the class of operators that can have a non-trivial v.e.v. in a SUSY vacuum.
However, the chiral operator on the LHS can be written as a SUSY
variation of another gauge invariant operator, and thus it trivially
has a vanishing v.e.v. in a SUSY vacuum. We are thus left with
a set of relations which can be solved for the v.e.v.s of $X_r$ in
terms of the couplings $g_r$ and $S$, the gluino condensate.
Once this is done, one can plug these values into the relation \refe{intin}
and obtain linear differential equations for the coupling dependent part of
$W_{eff}$, that is $W_{pert}$. The last step consists in adding the
relevant $S$-dependent (VY) superpotential, by considering some limit where the
low energy physics simplifies.

It is straightforward to apply this method to analyze, for instance,
$SU(N_c)$ SQCD with some flavors, $N_f<N_c$. By adding a mass term
for the quark superfields, one finds an expression for the meson
matrix in terms of the mass matrix and $S$, which then, after taking
into account the pure $SU(N_c)$ low energy physics, integrating out
the glueball superfield and integrating in the meson superfield,
leads to the celebrated Affleck-Dine-Seiberg superpotential:
\beq
W_{non-pert}= (N_c-N_f)\left({\Lambda^{3N_c-N_f} \over \det M}\right)^{1 \over
N_c-N_f}. \label{ads}
\eeq
Note that this method directly fixes the numerical coefficient, without
having to ultimately resort to instanton computations.

We now turn to a different set up, chiral theories
which have been argued to display dynamical SUSY breaking \cite{ads,mv}.
Some of these have been studied along these lines in \cite{us}.
Here we devote our attention to one of the simplest chiral theories, 
namely $SO(10)$ with one matter field $\Sigma$
in the spinorial $\bf 16$ representation.
This theory is peculiar because there are no invariants that can be written.
This means that there is no classical moduli space, but also no way
a superpotential, either tree level or non-perturbative, can be written.

One can however write a Konishi anomaly relation as in \refe{koni}:
\beq
\bar D^2 \Sigma^\dagger e^V \Sigma =  -  4 S. \label{trivial}
\eeq
In a SUSY vacuum, the consequence of \refe{trivial} would be
to impose $S=0$. However two lines of arguments can be brought up against
this conclusion. Following \cite{ads}, one could argue that in such
a symmetric vacuum, anomaly matching of unbroken $U(1)$ global symmetries
would imply a highly unlikely SUSY effective theory.
Or, in the spirit of \cite{mv}, a strong coupling instanton computation
would yield $\langle S \rangle \neq 0$ in the vacuum.

Both of these lines of arguments lead to removing the assumption
that there exists a SUSY vacuum. However since there is no small
parameter controlling the SUSY breaking, this is often referred to
as dynamical SUSY breaking at strong coupling. Another well known
and very similar 
example is the one of $SU(5)$ with matter in the antisymmetric $\bf 10$ and 
anti-fundamental $\bf \bar 5$ representations.

At this point we use a trick to make the theory ``calculable'', also 
employed in \cite{murayama} albeit in a different approach.
We add a flavor $v$ in the $\bf 10$ of $SO(10)$. Now we can write two
invariants:
\beq
X=v^2 \qquad \mbox{and} \qquad Y=v \Sigma^2,
\eeq
and we can use them to write a tree-level (renormalizable) superpotential:
\beq
W_{tree} = m X + \lambda Y.
\eeq
The Konishi anomaly relations now read:
\beqs
\bar D^2 v^\dagger e^V v &=&  2m X + \lambda Y -  2 S, \\
\bar D^2 \Sigma^\dagger e^V \Sigma &=& 2\lambda Y  -  4 S.
\eeqs
These relations now imply that in a SUSY vacuum $Y=2S/\lambda$ and,
if $m\neq 0$, $X=0$.

At this point we can either use the equations of motion of the fundamental
fields, or use a ``classical'' generalization of the Konishi relations
(where the one-loop term is vanishing) to argue that when $m\neq 0$,
we would also have $Y=0$ in a SUSY vacuum, and consequently also $S=0$.
To prevent being back in the same situation as before, we take for the time
being $m=0$. Thus classically $X$ can take any value and parametrizes
the vacuum. 
On the other hand the Konishi relations fix the v.e.v. of $Y$, so that
we can write:
\beq
Y= {2S\over \lambda} = {\partial W_{eff} \over \partial \lambda}.
\eeq
This is straightforwardly solved as:
\beq
W_{eff}= C(S) + S \log \lambda^2.
\eeq
To determine $C(S)$, we observe that when $X\neq 0$, the gauge symmetry
is broken to $SO(9)$. Moreover $\Sigma$ acquires a mass through the
tree level superpotential. Thus for large enough v.e.v. $X \gg \Lambda^2$,
the low energy theory is expected to be pure $SO(9)$ SYM.
The full effective superpotential is then:
\beq
W_{eff}= 7 S (1-\log {S\over \Lambda^3}) +  S \log \lambda^2.
\eeq
It implies for instance that there are 7 SUSY vacua, all of them additionally
parametrized by $X$.

We can now integrate in $Y$ and integrate out $S$ to get:
\beq
W_{non-pert} = 5 \left({4 \Lambda^{21} \over Y^2}\right)^{1\over 5}.
\label{wnp}
\eeq
This is a runaway superpotential like \refe{ads}.

Now we can consider turning on the mass term $mX$ 
in the tree-level superpotential. Since we know that already classically
this term is compatible with a SUSY vacuum only for $Y=0$, that is, 
precisely at the singularity of \refe{wnp}, we see that the extremization
of the full superpotential will have no solutions. However, since
the mass term lifts all the flat directions, we still expect the 
{\em potential}
of the theory to have a non-SUSY minimum, with the location being 
given in terms of $m$. For small $m$, we expect the minima to be located
at large values of the invariants $X$ and $Y$, which means that the 
original $SO(10)$ gauge symmetry is broken at weak coupling (of course here
after gauge symmetry breaking a non abelian gauge symmetry subsists,
which makes the low energy physics still strongly coupled, unlike
the ``calculable'' theory of \cite{ads}).
To recover the original model, one takes the limit $m\rightarrow \infty$,
so that the matter field $v$ decouples. This simply brings back the
SUSY breaking vacuum to a strong $SO(10)$ coupling, thus consolidating
(and systematizing) the arguments made before on the theory without
the vectorial matter. 

To conclude, in this contribution we hope to have convinced the reader
that the Konishi anomaly provides a systematic way to derive 
non-perturba\-ti\-ve superpotentials in purely ${\cal N}=1$ theories,
that is theories where the method \cite{cdsw}
of generalizing the Konishi anomaly is typically not available.

\subsection*{Acknowledgments}
It is a pleasure to thank Gabriele Ferretti and Rainer Heise for 
a fruitful collaboration on this and related work, and the organizers
of the Johns Hopkins Workshop in G\"oteborg and the RTN Workshop in
Copenhagen for giving me the opportunity to present it, and for the nice
atmosphere.
This work was and is partly supported by EU contracts HPRN-CT-2000-00122 
and HPRN-CT-00131, by the ``Actions de Recherche
Concert{\'e}es'' of the ``Direction de la Recherche Scientifique -
Communaut{\'e} Fran{\c c}aise de Belgique", by a ``P\^ole
d'Attraction Interuniversitaire" (Belgium) and by IISN-Belgium
(convention 4.4505.86). R.A. is a Postdoctoral Researcher of
the Fonds National de la Recherche Scientifique (Belgium).

\end{document}